\documentclass[pra,preprint,showpacs]{revtex4-2}
\usepackage[centertags]{amsmath}
\usepackage[utf8]{inputenc}
\usepackage{amsfonts}
\usepackage{amssymb}
\usepackage{amsthm}
\usepackage{amsmath}
\usepackage{newlfont}
\usepackage{epsfig}
\usepackage{amscd}
\usepackage{graphicx}
\usepackage{epsfig}
\usepackage{footnote}
\usepackage{lipsum}
\usepackage{color}
\usepackage{xcolor}
\usepackage{graphicx}
\usepackage{multirow}
\usepackage{enumerate}
\usepackage{caption}
\usepackage{subcaption}

\usepackage{amscd}

\newcommand{\beq}{\begin{equation}}
\newcommand{\eeq}{\end{equation}}
\newcommand{\ba}{\begin{array}}
\newcommand{\ea}{\end{array}}
\newcommand{\bea}{\begin{eqnarray}}
\newcommand{\eea}{\end{eqnarray}}
\begin{document}

\begin{center}
{\large \sc \bf {Measurement-based quantum state transfer and restoring via spin-1/2 chain interacting with environment  }
}

\vskip 15pt 

{\large
 E.B.~Fel'dman$^{1}$, 
 A.I.~Zenchuk$^{1,*}$
}

\vskip 8pt

{\it $^1$
Federal Research Center of Problems of Chemical Physics and Medicinal Chemistry RAS,
Chernogolovka, Moscow reg., 142432, Russia}.

{\it $^*$Corresponding author. E-mail:  zenchuk@itp.ac.ru}
\vskip 8pt

\end{center}

\begin{abstract}
We consider the multi-qubit fixed-excitation state transfer along the spin chain with dipole-dipole interaction subjected to the interaction  with environment  governed by the Lindblad equation preserving the excitation number during spin-evolution. The state transfer algorithm includes the state restoring  via Kraus operators  and ancilla measurement.  As a result, the transferred state appears in superposition  with completely mixed state, the latter  disappears with vanishing  interaction with environment. In that case we deal with probabilistic perfect state  transfer. 
Example of an arbitrary  multi-qubit one-excitation state transfer is present and its robustness with respect to perturbation of the  Kraus operators is studied.  
\end{abstract}


\maketitle

\section{Introduction}

The problem
 of an arbitrary state transfer, first formulated by Bose \cite{Bose}, is an intensively developing direction in quantum informatics. The phenomena of perfect \cite{CDEL,KS} and high-fidelity \cite{GKMT,GMT} state transfer were the first result of this study.  The  high-fidelity state transfer (HFST) turns out to be the most promising because it is more stable with respect to the Hamiltonian perturbations \cite{ZASO,ZASO2,ZAS_2015}. 
Among the methods of HFST, we highlight  the week-end-bond  method \cite{GKMT,GMT,FZ_2009,FKZ_2010,LMDMBA} and the method involving  the inhomogeneous magnetic field 
\cite{DZ_2010}.   In addition, the HFST controlled by the weak bonds  in  multi-dimensional spin arrays  governed by the XXZ-Hamiltonian  is considered in \cite{HAPM}. 
Set of papers is devoted to the state transfer along  chains interacting with  environment. Thus,  
 the high-fidelity state transfer along the spin-chain governed by the engineered  nearest-neighbor XX-Hamiltonian  with the Lindblad operators responsible  for interaction with environment  is studied in  Ref.\cite{XALW}.  
 The pulse-controlled state  transfer along a short homogeneous chain with non-Markovian dynamics is considered in Ref. \cite{WRLYW}. The   high-fidelity adiabatic state transfer based on  quantum metric  tensor is studied in \cite{VBR} and applied to both Hamiltonian and Lindblad evolution. The  mixed state transfer and restoring via controlled interaction with environment is studied in \cite{FLPZ__QIP_2026,FLPZ_RM_2026}.

In our paper,  we consider the problem of pure state transfer along the  spin-1/2 chain  under interaction with environment.  Our method is based on the state-restoring algorithm developed in Refs. \cite{FZ_2017,BFZ_2018,Z_2018,FPZ_2021,BFLP_2022,FPZ_2024}. The idea of this method is implementing the unitary transformation at the receiver side (at the extended receiver) to arrange proportionality between the elements of the sender density matrix and receiver density matrix which is achievable  for the nondiagonal matrix elements. It is important that the proportionality coefficients do not depend on a particular transferred state, i.e., they are universal  in that sense and defined by the Hamiltonian and time instant selected for state registration. The restoring algorithm  is applicable to transferring   multi-qubit states. 

In Ref.\cite{FWZ_arxive2025}, this method was combined with the ancilla-measurement  at the final step of state transfer with the purpose of removing the extra terms in the superposition state (garbage) and thus arranging the probabilistic perfect state transfer along the spin chain governed by the $XX$-Hamiltonian without interaction with environment.   
 
It is remarkable that  implementing  measurement into quantum algorithms is a promising  direction in quantum  information theory  \cite{USBGSN}.
Involving  measurement is multipurpose. First of all, measurement  can  extract the classical information  from the quantum system, for instance, 
in Shor's \cite{Shor1} and Grover's \cite{Grover} algorithms.
The measurement of the  objective function at fixed values of optimization parameters  in variational algorithms yields the input to the classical optimization algorithm  which calculates the subsequent values of optimization parameters  \cite{CABBEF,LFYYFZDM}.  
The measurement results in the teleportation algorithm \cite{Popescu,HHH},
  error mitigation algorithm  \cite{BGKMSZ}, HHL-algorithm for solving linear systems \cite{HHL}  are used  for further evaluation of the algorithm. 
The measurement is used in the algorithm constructing   the eigenvalues of non-unitary matrices  \cite{WWLN}.  
The measurement of certain (entangled) states can be  used in a special type of    quantum computation \cite{Wei},  in the quantum repeater creating remote entangled states  \cite{ZDB},
 in the long-distance  communication  \cite{ZBD},  
in the algorithm of state  preparation \cite{ZYY}, 
in the matrix-algebra  algorithms  \cite{ZQKW_2024, ZBQKW_arXive2024}.
  A large  class of quantum information problems  is covered  by the quantum  control, where measurement also plays a principal role. For instance,  the quantum control of dynamics of a two-level system  by  the non-selective von Neumann measurements of the system of optimal observables is studied in \cite{PISR}. Further development of this method gives rise to the  coherent and non-coherent control   of three-level \cite{SZPWSR,EP} and multi-level   \cite{SPHR}  quantum 
dynamics and the dynamics of open quantum systems \cite{WPRHT}. 

Here we apply the method implemented in Ref. \cite{FWZ_arxive2025} for perfect transfer of  pure states to transfer  the pure state under interaction with environment. 
In this case, the initial pure state becomes mixed one destroying the possibility of perfect state transfer (PST). We show that the transferred state can be restored up to the superposition with the maximally mixed state.  

The paper is organized as follows. 
The state restoring algorithm for Lindblad evolution is presented in Sec.\ref{Section:T} and based on the system of Kraus operators. Numerical example of two-qubit one-excitation state transfer is given in Sec.\ref{Section:num}, where the robustness of the restoring Kraus operators with respect to perturbations is demonstrated. The obtained  results are discussed in Sec. \ref{Section:conclusions}. Some details on analytical calculation of $\lambda$-parameters in the state transfer without interaction with environment are given in the Appendix, Sec.\ref{Section:appendix}.

\section{Algorithm for  state restoring}
\label{Section:T}
\subsection{State evolution and restoring}
\begin{figure*}[!]
\centering
\includegraphics[scale=0.65]{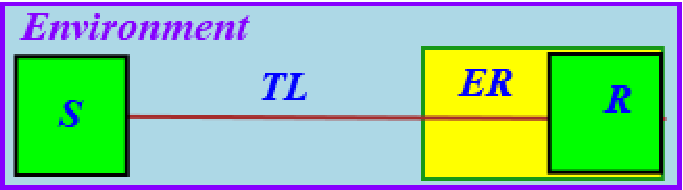}
\caption{Three-partite communication  line, consisting of the sender $S$, transmission line $TL$ and receiver $R$,   interacting with environment. The extended receiver $ER$ is also presented.}  
\label{Fig:circuit}
\end{figure*}

We consider the communication line, consisting of sender ($S$), receiver ($R$) and transmission line ($TL$) connecting them,  see Fig.\ref{Fig:circuit}.  
The number of spins in these subsystems is, respectively, $n^{(S)}$, $n^{(R)}$ and $n^{(TL)}$. 
All together, there are $N$ qubits in the communication line, $N= n^{(S)}+n^{(R)}+n^{(TL)}$. In addition, all spins of the communication line interact with environment.  We also consider the extended receiver \cite{FPZ_2021,BFLP_2022} which includes  the spins of receiver   $R$ together with several spins of $TL$. These several spins are  called $A$ (ancilla), and the transmission line without the subsystem $A$ is called $TL'$.  
Thus, $ER=R\cup A$, $TL=TL'\cup A$. 

Under interaction with environment, the state transfer   is governed by the Lindblad equation 
\begin{eqnarray}\label{L}
\partial_t \rho  = -i [H,\rho] + \sum_{j=1}^N \Gamma_j \left(  L_j\rho  L_j^\dagger - \frac{1}{2} L_j^\dagger L_j\rho - \frac{1}{2}  \rho L_j^\dagger L_j \right)
\end{eqnarray}
with the  pure initial state
\begin{eqnarray}
|\Psi(0)\rangle = |\psi(0)\rangle_S |0\rangle_{TL,R}.
\end{eqnarray}
It was shown in \cite{FWZ_arxive2025} that the $k$-excitation state transfer is simpler for realization than the  arbitrary state transfer provided that the excitation number is conserved during evolution. The algorithm for the  measurement-based perfect  transfer of the $k$-excitation state  along the spin chain governed by the Hamiltonian preserving the excitation number in the spin system was proposed therein. Therefore, we consider the transfer of the $k$-excitation initial pure sender state   $\displaystyle|\psi(0)\rangle_S = \sum_{j=0}^{N^{(S)}_k-1} s_j |\chi^{(k)}_j\rangle_S$, where 
$ |\chi^{(k)}_j\rangle_S$, $j=0,\dots, N^{(S)}_k-1$, represent the ordered basis in the $k$-excitation subspace.

Thus, in our paper, we consider the excitation preserving dynamics. In this case, both $H$ and $L_i$ have the block-diagonal structure in the basis ordered by the excitation number:
\begin{eqnarray}
H= \{H^{(0)},H^{(2)},\dots\},\;\; L_i = \{L_i^{(0)},L_i^{(1)},\dots\},
\end{eqnarray}
where $H^{(k)}$ and $L_i^{(k)}$ govern the $k$-excitation dynamics.

At the selected time instant for state registration $t_0$,  we can write for the elements of $\rho(t_0)$:
 \begin{eqnarray}
 \rho_{nm}(t_0) =\sum_{j=0}^{N^{(S)}_k} \sum_{l=0}^{N^{(S)}_k}  U_{nm;jl} \rho^{(S)}_{jl}(0),\;\; \rho^{(S)}(0) = |\psi(0)\rangle_S \; _S\langle \psi(0)|,\;\; \rho^{(S)}_{jl}(0) = s_j s_l^*,
 \end{eqnarray}
 where $*$ means the complex conjugate,  $U_{nm;jl}$ are the  elements of the operator $\hat U=e^{\hat u t}$, where the matrix $\hat u$ is determined by its elements
 \begin{eqnarray}
\hat u_{nm;ij} = - i(H_{ni} \delta_{jm} - \delta_{ni} H_{jm}) + \sum_{l=1}^N \Gamma_l \Big(Z_{l})_n (Z_l)_m - \frac{1}{4}\Big)  \delta_{ni}\delta_{jm},\;\;n,m,i,j =0, \dots N_k.
 \end{eqnarray}
 In $\hat u_{nm;ij}$, $nm$ and $ij$ are the double-indexes enumerating, respectively, rows and columns of the matrix $\hat u$. 
 Hereafter we use the indexes with subscripts indicating the particular subsystem whose state-space this index runs. We also use the multi-indexes (for instance, $i=(i_{S},i_{TL},i_R)$) to distinguish particular subsystems.
 At the selected  time instant $t_0$  after start of the evolution  governed by the master equation (\ref{L}), we apply the restoring  Kraus operators  $K^{(p)}$ to the extended receiver:
\begin{eqnarray}\label{rhoR}
&&\tilde{\rho}(t_0)=\sum_{p=1}^{N^{(K)}}(I_{S,TL'}\otimes K^{(p)}_{ER})\rho(t_0)(I_{S,TL'}\otimes K^{(p)}_{ER})^\dagger),\\\label{ort0}
&& \sum_{p=1}^{N^{(K)}}  (K^{(p)}_{ER})^\dagger  K^{(p)}_{ER} = I_{ER} ,
 \end{eqnarray}
 where  $N^{(K)}$ is the number of Kraus operators. In terms of matrix elements, Eq.(\ref{rhoR}) can be written as
 \begin{eqnarray}\label{rhoR2}
&&\tilde{\rho}_{nm}(t_0)=\sum_{p=1}^{N^{(K)}} K^{(p)}_{n_{ER};i_{ER}} 
\rho_{n_Sn_{TL'} i_{ER} ; m_Sm_{TL'} j_{ER}}(t_0) (K^{(p)}_{m_{ER};j_{ER}})^*)= T_{nm;i_S j_S} \rho^{(S)}_{i_S;j_S}(0),
 \end{eqnarray}
 where, $n=(n_{S},n_{TL'},n_{ER})$, $m=(m_{S},m_{TL'},m_{ER})$,
  \begin{eqnarray}\label{T}
&&T_{nm;i_S,j_S} = \sum_{i_{ER},j_{ER}}    W_{n_{ER} m_{ER};i_{ER}j_{ER}} 
 U_{(n_Sn_{TL'} i_{ER}) (m_S m_{TL'} j_{ER}) ;i_S,j_S},\;\; \\\label{WK}
 && W_{n_{ER} m_{ER};i_{ER}j_{ER}} =  \sum_p  K^{(p)}_{n_{ER};i_{ER}}  (K^{(p)}_{m_{ER};j_{ER}})^*
\end{eqnarray}
with the symmetry
\begin{eqnarray}
\label{symT}
&&T_{nm;i_S,j_S} =T^*_{mn;j_S,i_S}.
 \end{eqnarray}
We refer to the  operator 
\begin{eqnarray}\label{Wrestore}\
W_{ER}=\{ W_{n_{ER} m_{ER};i_{ER}j_{ER}} \}
\end{eqnarray}
 as  the  restoring transformation 
 In $\tilde \rho$,  we select the  $N^{(R)}_k\times N^{(R)}_k$ submatrix $\rho^{(R)}$, 
 \begin{eqnarray}\label{blockRho}
\tilde  \rho =\left( \begin{array}{cc}
 *&*\cr
 *&\rho^{(R)}
 \end{array} \right),
 \end{eqnarray}
 which eventually will be the receiver's density matrix (after proper normalization).
 Here stars mean the blocks of appropriate dimensions. For the elements of $\rho^{(R)}$, we have
 \begin{eqnarray}
&&\rho^{(R)}_{n_{R}m_{R}}(t_0)=\tilde \rho_{0_{S,TL},n_{R}; 0_{S,TL},m_{R}}(t_0) = T_{n_{R}m_{R};i_S,j_S}\rho^{(S)}_{i_Sj_S}(0),\\\nonumber
&&
 T_{n_{R}m_{R};i_S,j_S} \equiv  T_{(0_S0_{TL}n_{R})(0_S0_{TL}m_{R});i_S,j_S}
.
 \end{eqnarray}
 We recall that the element  $\rho^{(R)}_{ij}$ is restored if it is proportional to the appropriate element of the sender's initial state $\rho^{(S)}_{ij}$ \cite{FPZ_2021}. This requirement  creates certain constraints for the elements of $T$ which form the restoring system. 
The restoring system  for the nondiagonal elements is given by 
\begin{eqnarray} \label{rest1}
&&
 T_{n_{R}m_{R};i_S,j_S} = \lambda_{i_{S}j_{S}} \delta_{n_{R} i_{S}} \delta_{m_{R} j_{S}} ,  \;\;  n_{R} \neq m_{R} \;\; \Leftrightarrow \;\; 
 \\\label{rest11}
  &&T_{n_{R}m_{R};i_S,j_S} = 0, n_{R} \neq m_{R}, \;\; (n_{R},  m_{R} ) \neq (i_{S} ,j_{S}), \\\label{rest12}
  &&
  T_{i_{S}j_{S};i_S,j_S} = \lambda_{i_{S}j_{S}},\;\; i_S\neq j_S.
  \end{eqnarray}
 In addition, we  require that all $\lambda$-parameters are positive  and equal to each other:
  \begin{eqnarray}
  \label{rest13}
 &&  \lambda_{i_{S}j_{S}} = T_{i_{S}j_{S};i_S,j_S}=\lambda>0 ,  \;\;  i_{S} \neq j_{S}\;\;\Rightarrow \;\;  {\mbox{Im}} \, T_{i_{S}j_{S};i_S,j_S} =0.
 \end{eqnarray}
 For the diagonal elements of $\rho^{(R)}$ we set
 \begin{eqnarray}
  \label{rest2}
 && T_{n_{R}n_{R};i_S,j_S}  =0,\;\; i_S\neq j_S,\\\label{rest3}
 && T_{n_{R}n_{R};n_R,n_R}  =\lambda + \nu,\\\label{rest4}
 && T_{n_{R}n_{R};j_S,j_S}  =\nu,\;\; n_R\neq j_S,
\end{eqnarray}
where $\lambda$ and $\nu$ are real  non-negative parameters. 
In this case, the block $\rho^{(R)}$ in the density matrix (\ref{blockRho})  reads
\begin{eqnarray}\label{protr}
\rho^{(R)} (t_0)= \nu I_R + \lambda \rho^{(S)}(0),
\end{eqnarray}
where $I_R$ is the identity $N^{(R)}_k\times N^{(R)}_k$ operator. We note that $\rho^{(R)} (t_0)$ is not a density matrix because ${\mbox{Tr}}\,\rho^{(R)} (t_0)<1$.

The number of free parameters in the Kraus operators must be enough to solve all restoring constraints (\ref{rest11}), (\ref{rest13})-(\ref{rest4}). To determine the minimal required dimension of the extended receiver we compare the number of free parameters in the Kraus operators and the number of restoring constraints. 

 Let us calculate the number of free real parameters first.  Each Kraus operator  yields  
 $2 (N^{(ER)}_k)^2$ real parameters. Thus, we have 
$ 2(N^{(ER)}_k)^2 N^{(K)}$  parameters in $N^{(K)}$  Kraus operators. 
These parameters are subjected to orthogonality condition  (\ref{ort0}) which can be conveniently written as
 \begin{eqnarray}\label{ort12}
 \sum_{p=1}^{N^{(K)}} \Big((K^{(p)}_{ER})^\dagger  K^{(p)}_{ER}\Big)_{i_{ER}j_{ER}} =\delta_{i_{ER}j_{ER}},\;\; i_{ER},j_{ER} =0 ,\dots, N^{(ER)}_k-1. 
\end{eqnarray}
There are  $(N^{(ER)}_k)^2$ real equations in  (\ref{ort12}). Therefore, we have 
$N^{(par)} =2(N^{(ER)}_k)^2 N^{(K)}  -(N^{(ER)}_k)^2=     (N^{(ER)}_k)^2 ( 2  N^{(K)}- 1)$ free real parameters. 

Now we calculate the  number of restoring constraints. 
The number of real constraints  in (\ref{rest11}) (in view of symmetry (\ref{symT}))  is $N^{(R)}_k (N^{(R)}_k-1)((N^{(R)}_k)^2 -1) $. The number of real equations in (\ref{rest13}) (in view of symmetry (\ref{symT})) is $N^{(R)}_k ( N^{(R)}_k-1) -1$.  The numbers of real equations in (\ref{rest2}), (\ref{rest3}) and (\ref{rest4})  are, respectively,  $(N^{(R)}_k)^2 ( N^{(R)}_k-1)$,  $N^{(R)}_k $ and 
$N^{(R)}_k(N^{(R)}_k-1)-1$.  All together, we have $N^{(eq)}=(N^{(R)}_k)^4- 2$ real conditions.  To satisfy $N^{(eq)}$ constraints using $N^{(par)}$ free parameters,  we require $N^{(par)}\ge N^{(eq)}$, i.e.,
\begin{eqnarray}
\label{pareq}
(N^{(ER)}_k)^2 \ge \frac{1}{2N^{(K)}-1}\left( (N^{(R)}_k)^4- 2  \right).
\end{eqnarray}
 
 We note that condition (\ref{pareq}) is  satisfied for  $N^{(ER)}_k=N^{(R)}_k$, $N^{(K)} = (N^{(R)}_k)^2$ at any $N^{(R)}_k>0$, therefore this condition does not create the strong  lower boundary for the dimensionality of the extended receiver $N^{(ER)}$ which can coincide with $N^{(R)}$ to satisfy  inequality (\ref{pareq}). 
However,  the lower boundary for $N^{(ER)}$ is fixed by another constraint, which is formulated in the following Proposition.  

{\bf Proposition.} 
The minimal dimensionality of the extended receiver $n^{(ER)}$ is defined by the inequality
\begin{eqnarray}
\label{NER}
N^{(ER)}_k \ge N^{(ER;min)}_k \equiv N^{(R)}_k+1\;\;\;\Rightarrow \;\;\; n^{(ER)} \ge n^{(ER;min)} \equiv n^{(R)}+1.
\end{eqnarray}

{\it Proof.} Let $a_{n_{ER}m_{ER}}$ be the row of the elements 
$ W_{n_{ER} m_{ER};i_{ER}j_{ER}}$, $ i_{ER}, j_{ER} = 0,\dots, N^{(ER)}_k-1$, and $b_{i_S j_S}$ be the column of the elements 
 $U_{i_{ER}j_{ER};i_S,j_S}$,  $ i_{ER}, j_{ER} = 0,\dots, N^{(ER)}_k-1$, 
 i.e.,
 $T_{n_{ER}m_{ER};i_Sj_S} = a_{n_{ER}m_{ER}}  b_{i_Sj_S}$.  Here we use double indexes to enumerate the rows and columns of the matrices. Then the restoring conditions (\ref{rest1}), (\ref{rest3}) can be written as
 \begin{eqnarray}\label{RC}
 a_{n_S n_S} b_{i_S j_S} = \delta_{(n_S n_S)(i_S j_S)} (\nu \delta_{i_S j_S} +\lambda),  
 \end{eqnarray}
 where 
 $$
 \delta_{(n_S n_S)(i_S j_S)} =\left\{
 \begin{array}{ll}
 1& {\mbox{if}}\;\; n_S=i_S \;\;{\mbox{and}}\;\; m_S=j_S\cr
 0& {\mbox{otherwise}}
 \end{array}
 \right.
 .
 $$
 Eq.(\ref{RC})  requires
 \begin{eqnarray}\label{bSS}
 &&b_{i_S j_S} =  (\nu \delta_{i_S j_S} +\lambda) \tilde a_{i_S j_S} + \\\nonumber
 &&
 \left(\sum_{i=0}^{N^{(ER)}_k-1} \sum_{j=N^{(R)}_k}^{N^{(ER)}_k-1} + \sum_{i=N^{(R)}_k}^{N^{(ER)}_k-1} \sum_{j=0}^{N^{(ER)}_k-1} - \sum_{i=N^{(R)}_k}^{N^{(ER)}_k-1} \sum_{j=N^{(R)}_k}^{N^{(ER)}_k-1} \right)\alpha_{i_Sj_S;i j}  \tilde a_{ij}  .
 \end{eqnarray}
Here $\tilde a_{i_S j_S}$ are the columns of the matrix inverse to $W \equiv \{W_{n_{ER} m_{ER};i_{ER}j_{ER}}\}$ and $\alpha_{i_Sj_S;i j}$ are some constants.  Obviously, if $N^{(ER)}_k=N^{(R)}_k$, then the sums in the right hand side of  (\ref{bSS}) yield zero, consequently  the columns  $b_{i_S j_S}$ are proportional to the appropriate columns $ \tilde a_{i_S j_S}$  in this case and therefore they are orthogonal columns. However, these orthogonality conditions  would  be additional constraints on the elements of $U$ which are  not presumed by the evolution Hamiltonian. 

On the contrary, if condition (\ref{NER})  is satisfied, i.e., $N^{(ER)}_k>N^{(R)}_k$, then the sums in the right hand side of   Eq.(\ref{bSS}) are non-zero and additional constraints for the elements of the matrix $U\equiv \{U_{i_{ER}j_{ER};i_S,j_S}\}$   do not appear. In particular, following the strategy  of Refs.\cite{BFLP_2022,FWZ_arxive2025},  multiplying  Eq.(\ref{bSS}) by $b^\dagger_{n_Sm_S}$ from the left  does not generate any extra  constraint on the elements of the operator $U$, because $W$  is not a unitary operator in our case.
$\Box$

\subsection{Additional constraints maximizing $\lambda$}

To simplify maximization of $\lambda$, we  introduce some extra constraints  on the elements of the matrix $T = \{T_{nm;ij}\}$.  The purpose is to make as many zero elements in $\rho$ as possible.

For that, we add  two sets of constraints. The first one is represented by  the system  of equations (recall that the receiver is the last subsystem of the communication line)
\begin{eqnarray}\label{addconstr1}
 &&T_{n_{ER}m_{ER};i_S,j_S}  =0,\;\;
 i_S,j_S =0,\dots, N^{(S)}_k-1,\;\; (n_{ER},i_S)\neq (m_{ER},j_S),
  \\\nonumber
 &&
 n_{ER}=0,\dots, N^{(ER)}_k-2 ,\;\; m_{ER}=N^{(R)}_k,\dots, N^{(ER)}_k -1,
\end{eqnarray}
and analogous equations following from  symmetry (\ref{symT}).
In  (\ref{addconstr1}), the condition  $(n_{ER},i_S)\neq (m_{ER},j_S)$ follows from the property    $T_{n_{ER}n_{ER};i_S,i_S}>0$, which, in turn,  follows from the structure of the operator  $T$ given in Eqs. (\ref{T})- (\ref{symT}).

The second set of constraints is given by 
\begin{eqnarray}\label{addconstr2}
 &&T_{n_{ER}m;i_S,j_S}  =0,\;\; \\\nonumber
 &&
 i_S,j_S =0,\dots, N^{(S)}_k-1,\;\;n_{ER}= 0,\dots, N^{(ER)}_k-1,\\\nonumber
 &&
 m=0,\dots,N^{(S,TL,R)}_k-N^{(ER)}_k -  1.
\end{eqnarray}
and analogous equations following from symmetry (\ref{symT}).
Each of  systems  (\ref{addconstr1})  and  (\ref{addconstr2})  put zero appropriate elements of the density matrix $\rho(t_0)$.

 \subsection{Selecting the transferred state out of the density matrix $\rho(t_0)$}
 \label{Section:select}
\begin{figure*}[!]
\centering
\includegraphics[scale=0.60]{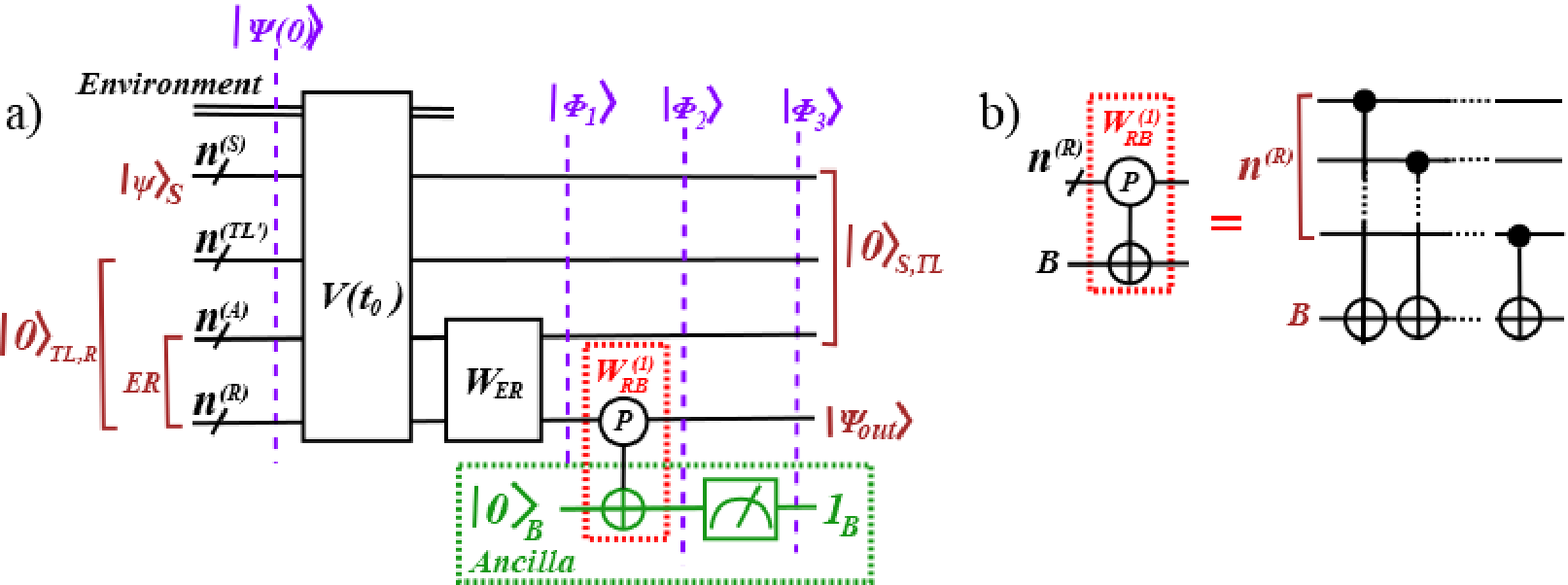}
\caption{ $k$-excitation state transfer and restoring under Lindblad equation (\ref{L}); (a) the general circuit; (b) the realization of $W^{(1)}_{RB}$ for the 1-excitation state transfer.}  
\label{Fig:circuit2}
\end{figure*}

 The density matrix at the time instant  $t_0$ after applying the restoring transformation can be written as 
  \begin{eqnarray}\label{rhog}
\tilde \rho =\left( \begin{array}{cc}
 0_{N^{(S)}_k + N^{(TL)}_k, N^{(S)}_k + N^{(TL)}_k}&0_{N^{(S)}_k + N^{(TL)}_k,N^{(R)}_k}\cr
0_{N^{(R)}_k,N^{(S)}_k + N^{(TL)}_k}&\rho^{(R)}
 \end{array} \right)+  \rho_g,\;\;\rho_g=\left( \begin{array}{cc}
 *&*\cr
 *&0_{N^{(R)}_k,N^{(R)}_k}
 \end{array} \right),
 \end{eqnarray}
 where $\rho^{(R)}$ is given in (\ref{protr}) and  stars are some non-zero blocks of appropriate dimensions. 
Our purpose is to  select the $k$-excitation block $\rho^{(R)}$. 
 To this end, we  introduce the projector
 \begin{eqnarray}
 P^{(k)}_R = \sum_{j=0}^{N^{(R)}_k-1} |\chi^{(k)}_j\rangle \, \langle \chi^{(k)}_j| 
 \end{eqnarray}
 and the one-qubit ancilla $B$ in the ground state. 
 Based on this projector, we prepare the controlled operator
 \begin{eqnarray}\label{Wk}
 W^{(k)} _{RB} = P^{(k)}_R\otimes \sigma^{(x)}_{B}+ (I_R- P^{(k)}_R)\otimes I_B.
 \end{eqnarray}
 Applying the  operator $W^{(k)} _{RB}$ to $\tilde \rho\otimes |0\rangle_B\, {_B\langle 0|}$
 we obtain 
 \begin{eqnarray}
 &&
 \rho^{(1)} = W^{(k)} _{RB} \Big(\tilde \rho\otimes |0\rangle_B\, {_B\langle 0|}\Big)   (W^{(k)} _{RB})^\dagger =\\\nonumber
&&
\rho^{(R)} \otimes |0\rangle_{S,TL}\; {_{S,TL}\langle 0|}\otimes |1\rangle_{B}  \, {_{B}\langle 1|} + \tilde \rho_g,\\\label{rhoRf}
 &&
\rho^{(R)}= 
  \sum_{i=0}^{N^{(R)}_k-1}\sum_{j=0}^{N^{(R)}_k-1}   
  r^{(kk)}_{ij}  |\chi^{(k)}_i\rangle \, \langle \chi^{(k)}_j| \stackrel{(\ref{protr})}{=}\nu I_R + \lambda \rho^{(S)}(0),
 \end{eqnarray}
where $\tilde \rho_g$ collects all extra elements of the density matrix $\rho^{(1)}$.  
 Finally, we remove  the garbage $\rho^{(R)}_g$ via applying the  projector $|1\rangle_B \, _B\langle 1|$ which is equivalent to   the measurement of $B$ with the output $|1\rangle_B$. The  success probability of such measurement is 
 \begin{eqnarray}\label{p}
 p=\lambda + N^{(R)}_k \nu.
 \end{eqnarray}
  As the result, we  obtain the state
  \begin{eqnarray}
  \rho^{(R)}_{out} =\frac{1}{p}\Big( \nu I_R + \lambda \rho^{(S)}(0)\Big),
  \end{eqnarray}
  which is the superposition of the sender state $\rho^{(S)}(0)$ and the  maximally mixed state $I_R/N^{(R)}_k$.
 The circuit for the described state transfer algorithm is presented in Fig.\ref{Fig:circuit2}.

\subsection{Realization of Kraus operators}

The Kraus operators can be realized in terms of  the trace of  a unitary  operator applied to a larger system.
Let us include the  $n^{(E)}$-qubit subsystem $E$ (which does not participate in the evolution) and extend   the density matrix $\rho$ attaching the $2^{n^{(E)}}\times 2^{n^{(E)}}$ maximally mixed density  matrix, whose dimensionality will be clarified below:
$\hat \rho = 1/2^{n^{(E)}}\rho\otimes I^{(E)}$.
Now we apply the unitary operator $V_{ER,E}$ (preserving the excitation number) to the subsystem $ER\cup E$  and calculate the partial trace over the qubits of $E$:
\begin{eqnarray}
&&
\tilde \rho = \frac{1}{2^{n^{(E)}}}
{\mbox{Tr}}_E  (I_{S,TL'}\otimes V_{ER,E}) (\rho\otimes I^{(E)}) (I_{S,TL'}\otimes V_{ER,E}^\dagger),
\end{eqnarray}
or, for the elements of $\tilde \rho$, 
\begin{eqnarray}\label{trho}
&&
\tilde \rho_{nm}  = \sum_{n_E=0}^{2^{n^{(E)}}-1} \sum_{i_E=0}^{2^{n^{(E)}}-1}  \sum_{i_{ER},j_{ER}=0}^{N^{(ER)}_k-1}   V_{n_{ER}n_E;  i_{ER} i_E} \rho_{n_Sn_{TL'} i_{ER};m_S m_{TL'}  j_{ER}}  V^\dagger_{j_{ER}i_E;m_{ER} n_E}=\\\nonumber
&&
 \sum_{n_E=0}^{2^{n^{(E)}}-1} \sum_{i_E=0}^{2^{n^{(E)}}-1} K^{(n_E ,i_E)}_{n_{ER};i_{ER}}  \rho_{n_Sn_{TL'} i_{ER};m_S m_{TL'}  j_{ER}}      (K^{(n_E ,i_E)})^\dagger_{j_{ER},m_{ER}},
\end{eqnarray}
where  $n=\{n_S,n_{TL'},n_{ER}\}$,  $m=\{m_S,m_{TL'},m_{ER}\}$,
$K^{(n_{E} ,i_{E})}_{n_{ER};i_{ER}} =V_{n_{ER}n_{E}; i_{ER} i_{E}} $, $V_{ER,E}=\{ V_{n_{ER}n_{E}; i_{ER} i_{E}}  \}$.
 Introducing the single index $p$ for  $(n_E ,i_E)$ we rewrite  (\ref{trho})  as
 \begin{eqnarray}
 \tilde \rho =\sum_{p=1}^{N^{(K)}} (I_{S,TL'} \otimes  K^{(p)}_{ER}) \rho  (I_{S,TL'} \otimes (K^{(p)}_{ER})^\dagger),\;\; N^{(K)}=2^{2n^{(E)}},
 \end{eqnarray}
 which coincides with (\ref{rhoR}) at the time instant $t_0$. 
Here $K^{(p)}_{ER} = \{ K^{(p)}_{n_{ER};m_{ER}}:  n_{ER},m_{ER} = 0,\dots, N^{(ER)}_k-1 \}$.
 The normalization (\ref{ort12}) for the Krause operators follows from the unitarity of $V_{ER,E}$.
 Thus, the parameters of the Kraus operators are the parameters of the unitary transformation $V_{ER,E}$ in the extended space.

\section{Numerical examples: one-excitation state transfer}
\label{Section:num}
We consider the one-excitation two-qubit state transfer with three-qubit extended receiver $ER$, in this case $N^{(R)}_1=N^{(S)}_1= 2$,
$N^{(ER)}_1=3$.  We set $\Gamma_j=\Gamma$ for all $j$ and introduce the  normalized parameter $\Lambda=\lambda/p$, $p$ is defined in (\ref{p}).
If we use the whole system (\ref{rest11}), (\ref{rest13})-(\ref{rest4}), (\ref{addconstr1}),  (\ref{addconstr2}) for restoring, then we denote the  parameters $\Lambda$ and $p$  by $\Lambda_1$ and $p_1$. If the restoring  system is (\ref{rest11}), (\ref{rest13})-(\ref{rest4}), (\ref{addconstr1}), then we use the notations $\Lambda_2$ and $p_2$.  Finally, if  system (\ref{rest11}), (\ref{rest13})-(\ref{rest4}) is used for restoring, then we denote the parameters as  $\Lambda_3$ and $p_3$. In all three above cases we use orthogonality condition (\ref{ort12}). Note that if $\Gamma=0$, then $\nu=0$ so that $\lambda=p$,  $\Lambda_i =1$, $i=1,2,3$,  $p_2=p_3=\lambda_2=\lambda_3$, because in this case system  (\ref{addconstr2}) becomes consequence of  system (\ref{addconstr1}).

For  numerical simulations,  we use the Lindblad  Eq.(\ref{L}) and the $XX$-Hamiltonian with all-node dipole-dipole interactions and external magnetic field directed perpendicular to the line chain:
\begin{eqnarray}\label{Hamiltonian}
H=\sum_{j>i}D_{ij} (I_{xi}I_{xj} +I_{yi}I_{yj} ), \;\;[H,I_z]=0,
\end{eqnarray}
where $I_{\alpha i}=\frac{1}{2}\sigma^{(\alpha)}$ is the operator of 
the $\alpha$-projection of the $i$th spin momentum, $\alpha=x, y, z$, $I_z=\sum_i I_{zi}$,  $\sigma^{(\alpha)}$  are the Pauli matrices,
$D_{ij}= \gamma^2\hbar/(2r_{ij}^3)$ is the dipole-dipole coupling constant between the $i$th and $j$th spins,  $r_{ij}$ is the distance between the $i$th and $j$th spins, 
$\gamma$ is the gyromagnetic ratio,  $\hbar$ is the Planck constant  ($\hbar =1$ for simplicity). In simulations, we usually use the homogeneous chains (so that all nearest-neighbor coupling constants equal to each other) and  the {  dimensionless time $\tau = t D_{12}$.}

\begin{figure*}[!]
\centering
\includegraphics[scale=0.65]{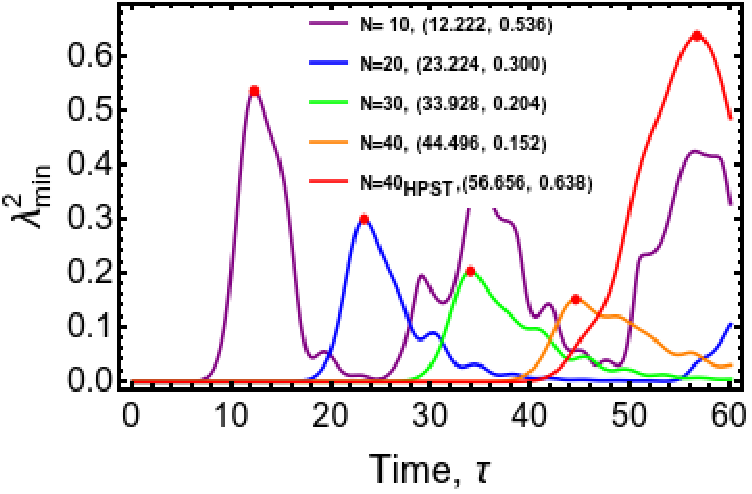}
\caption{Under neglecting the interaction with environment ($\Gamma=0$), the evolution of $\lambda$-parameters (minimal of two roots (\ref{roots})) for the homogeneous  chains of $N=10, 20, 30, 40$ nodes and for the HFST-chain of 40 nodes  is shown. Coordinates of the first maxima  are indicated in the legend and marked by the red circles on the graphs. }  
\label{Fig:Perfect}
\end{figure*}

Our purpose is to construct such restoring transformation  $W_{ER}$ (\ref{Wrestore})  that minimizes the ratio   $\nu/\lambda$ of the parameters  in Eq.(\ref{protr}). To fix the time instance for state registration $t_0$, we consider the evolution without interaction with environment, when  $\Gamma=0$ and $\nu=0$, and find the time instance maximizing $\lambda$. This can be done analytically following the protocol in Ref.\cite{FWZ_arxive2025}, see also Appendix for our case of the one-excitation two-qubit state transfer. 
The evolutions of $\lambda$-parameter for homogeneous chains of $N = 10, 20, 30, 40$ nodes  and for the 
HFST-chain \cite{ABCVV,FWZ_arxive2025}  of 40 nodes (with adjusted two  end-pairs of coupling constants)  are shown in Fig.\ref{Fig:Perfect} with selected time instants $t_0$ marked by the red circles. 

The first set of numerical experiments is performed for the chain of $N=10$ nodes using restoring systems (\ref{rest11}), (\ref{rest13})-(\ref{rest4}), (\ref{addconstr1}),  (\ref{addconstr2})  (the parameters $\Lambda_1$ and $p_1$ will be obtained ) and 
 (\ref{rest11}), (\ref{rest13})-(\ref{rest4})     (obtaining the parameters $\Lambda_3$ and $p_3$) with orthonormalization condition (\ref{ort12}) and $N^{(K)}=(N^{(ER)}_1+1)^2=16$ for both cases.
The  interaction strength is characterized by the following  set of values for the  dimensionless parameter $\tilde \Gamma =\Gamma/D_{12} $:
\begin{eqnarray}\label{gamma}
\tilde \Gamma = 0.0001, 0.001, 0.01, 0.1.
\end{eqnarray} 
We obtain 3000 independent solutions of the appropriate  restoring system and choose the solution which minimizes the ratio 
$\nu/\lambda$ adding the  equation 
\begin{eqnarray}\label{nulam}
\frac{\nu}{\lambda}=\gamma_0 a,
\end{eqnarray}
 with unknown $a$, $0\le a\le1$,  and $\gamma_0 =0.0007, 0.006, 1, 1.35$  for the respective  $\tilde \Gamma$ from  list (\ref{gamma}). 
 The optimized parameters $\Lambda_1$, $\Lambda_3$, $p_1$, $p_3$   are collected in Table \ref{Table:N10}. The fact that $p_3>p_1$ for $\tilde \Gamma =0.1$ is not contradictory because  we minimize the ratio $\displaystyle \frac{\nu}{\lambda}=\frac{1}{N^{(ER)}_k} \left(\frac{1}{\Lambda} -1\right)$ without paying attention to minimization of $p$.
 \begin{table}
\begin{tabular}{|c|ccccc|}
\hline
$\tilde \Gamma$ &0 &0.0001 &0.001 &0.01 &$0.1$\cr
\hline
$\Lambda_1$&1& 0.999 &0.991&0.748&0.375\cr 
$\Lambda_3$&1& 0.999& 0.991& 0.587& 0.305\cr 
$p_1$&0.536&0.527 & 0.525&0.414&0.211\cr 
$p_3$&0.405&0.534& 0.531& 0.306 &0.292\cr
\hline
\end{tabular}
\caption{Normalized $\lambda$-parameters $\Lambda_1$, $\Lambda_3$ and appropriate success probabilities $p_1$, $p_3$ for the 10-node spin chain with different strength of interaction with the environment $\tilde\Gamma$. The time instant for state registration is $\tau_0=12.222$.}
\label{Table:N10}
\end{table}

The second set of experiments is performed for the homogeneous  chains of different lengths
\begin{eqnarray}\label{long}
N=10, 20, 30, 40 ,40_{HFST},
\end{eqnarray}
where $40_{HFST}$ means the   HFST-chain of 40 nodes
with 
$D_{12} = D_{39,40}=0.510\delta $  and  $D_{23} = D_{38,39}=0.348\delta $, all other $D_{i,i+1}=\delta$ (in this case, the dimensionless parameters are  $\tau= \delta t$, $\tilde \Gamma=\Gamma/\delta$).  We use the restoring system  (\ref{rest11}), (\ref{rest13})-(\ref{rest4}), (\ref{addconstr1})    and  admit  $N^{(K)}=(N^{(ER)}_1)^2+3 = 12$. 
In this case $\gamma_0=1, 1.11, 1.19, 1.26, 0.125$ in Eq.(\ref{nulam}) for the chains of lengths (\ref{long}) 
 respectively. 
The parameters  $\Lambda_2$, $p_2$ and $\tau_0$  for the chains of  different length $N$ 
are collected in the Table \ref{Table:PST}. 
\begin{table}
\begin{tabular}{|c|ccccc|}
\hline
$N$ &10 &20 &30 &40 &$40_{HFST}$\cr
\hline
$\tau_0$&12.222&23.224&33.928&44.496&56.656\cr
$\Lambda_2|_{\tilde\Gamma=0}$&0.536&0.300&0.204&0.152&0.638\cr 
$\Lambda_2|_{\tilde \Gamma=0.01}$&0.679& 0.426&0.405&0.408&0.811\cr 
$p_2|_{\tilde\Gamma=0.01}$&0.394& 0.239&0.121& 0.102&0.397\cr 
\hline
\end{tabular}
\caption{The time-instants $\tau_0$,  parameter $\Lambda_2$ at $\tilde \Gamma=0, 0.01$ and success probability $p_2$ at $\tilde\Gamma=0.01$  for the state transfer  along the chains of different lengths $N$.}
\label{Table:PST}
\end{table}
This table demonstrates  that $p$ decreases with an increase in $N$ unless the HFST-chain is considered. In the latter case, the success probability for the 40-spin chain is even higher than for the 10-spin chain.

\subsection{Robustness with respect to perturbations of restoring Kraus operators}
Let the Kraus operators be subjected to perturbation with the amplitude $\varepsilon \ll 1$:
\begin{eqnarray}
\tilde K^{(p)} = K^{(p)} + \varepsilon {\cal{K}}^{(p)},
\end{eqnarray}
where we omit the subscript $ER$ at the operators $K^{(p)}$.
The perturbations  ${\cal{K}}^{(p)}$ are arbitrary up to the properly perturbed normalization condition (\ref{ort12}) which now is given by
\begin{eqnarray}\label{norm2}
 \sum_{p=1}^{N^{(K)}} \Big((\tilde K^{(p)})^\dagger  \tilde K^{(p)}\Big)_{i_{ER}j_{ER}} =\delta_{i_{ER}j_{ER}},\;\; i_{ER},j_{ER} =0 ,\dots, N^{(ER)}_k-1 , 
\end{eqnarray}
or, in view of (\ref{ort12}), 
\begin{eqnarray}\label{norm22}
 \sum_{p=1}^{N^{(K)}} \left( \Big((K^{(p)})^\dagger  {\cal{K}}^{(p)}\Big)_{i_{ER}j_{ER}} + \Big(({\cal{K}}^{(p)})^\dagger  {{K}}^{(p)}\Big)_{i_{ER}j_{ER}}    +\varepsilon
  \Big(({\cal{K}}^{(p)})^\dagger  {\cal{K}}^{(p)}\Big)_{i_{ER}j_{ER}} \right)  =0.
\end{eqnarray}
To modulate the random perturbations ${\cal{K}}^{(p)}$ and  satisfy condition (\ref{norm22}), we consider the following form of the elements of ${\cal{K}}^{(p)}$:
${\cal{K}}^{(p)}_{ij}= r^{(p)}_{ij} e^{2 \pi i q^{(p)}_{ij}} {\cal{K}}_{ij}$, where  $r_{ij}$ and $q_{ij}$ are the random real numbers, $0\le r^{(p)}_{ij}\le 1$,  $0\le q^{(p)}_{ij}\le 1$, and ${\cal{K}}_{ij}$ are roots of the normalization system (\ref{norm22}).

Under perturbation, the operator $T$ in the restoring condition (\ref{rest1}) becomes perturbed operator $\tilde T$, the value of this perturbation can be  estimated  by the 
quantity 
\begin{eqnarray}\label{deltaT}
\delta^{(T)} =\sqrt{\frac{\displaystyle\sum_{n_R,m_R;i_S,j_S}  |T_{n_R,m_R;i_S,j_S} - \tilde T_{n_R,m_R;i_S,j_S}|^2}{\displaystyle\sum_{n_R,m_R;i_S,j_S}  |T_{n_R,m_R;i_S,j_S}|^2}}.
\end{eqnarray}
We also introduce the quantity
\begin{eqnarray}\label{deltaK}
\delta^{(K)} =  \sqrt{\frac{ \displaystyle\sum_{p=1}^{N^{(K)}} \sum_{n_{ER},m_{ER},i_{ER},j_{ER}}  |{\cal{K}}^{(p)}_{n_{ER},m_{ER};i_{ER},j_{ER}}|^2}{\displaystyle \sum_{p=1}^{N^{(K)}} \sum_{n_{ER},m_{ER},_{ER},j_{ER}}  |{{K}}^{(p)}_{n_{ER},m_{ER};i_{ER},j_{ER}}|^2}}
\end{eqnarray}
to estimate the  perturbation influence on  the system of Kraus operators.

\begin{figure*}[!]
\centering
\includegraphics[scale=0.65]{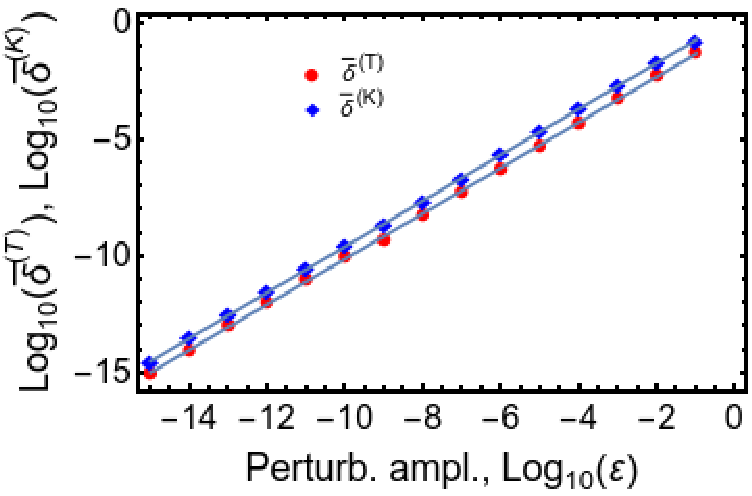}
\caption{Dependence of parameters $\overline{\delta}^{(T)}$ and $\overline{\delta}^{(K)}$ on $\varepsilon$ for $N=10$ and $\tilde\Gamma=0.0001$.}  
\label{Fig:T}
\end{figure*}

For each value of $\varepsilon$ from the  set $\varepsilon = 10^{-j}$, $j=1,\dots,15$, we average $\delta^{(T)}$ and $\delta^{(K)}$ over  1000 random perturbations and denote the averaged values as $\overline{\delta}^{(T)}$ and 
$\overline{\delta}^{(K)}$. The results for the first set of experiments with $N=10$ and  restoring system  (\ref{rest11}), (\ref{rest13})-(\ref{rest4}), (\ref{addconstr1}),  (\ref{addconstr2})  are shown in Fig.\ref{Fig:T} for $\tilde\Gamma=0.0001$ by circles and diamond respectively. Both sets of points can be approximated by the linear functions $\log_{10} \overline{\delta}^{(T)} = 0.996   \log_{10} \varepsilon -0.030 $ and  
$\log_{10} \overline{\delta}^{(K)} = 0.995   \log_{10} \varepsilon  +0.306 $  as shown in Fig.\ref{Fig:T}. 
For other $\tilde\Gamma$ from list (\ref{gamma}) ($N=10$)
the values of the parameters  $\overline{\delta}^{(T)}$ and $\overline{\delta}^{(K)}$ belong to  the (almost) parallel lines  
\begin{eqnarray}\label{lines}
\log_{10} \overline{\delta}^{(T)} = \varkappa^{(T)}   \log_{10} \varepsilon  +C^{(T)} , \;\;\; \log _{10}
\overline{\delta}^{(K)} = \varkappa^{(K)}   \log_{10} \varepsilon  +C^{(K)} 
\end{eqnarray}
 with the parameters $\varkappa^{(T)}$, $C^{(T)}$ and $\varkappa^{(K)}$, $C^{(K)}$ collected in Table \ref{Table:3}a.
It is interesting, that the similar perturbation analysis performed for the chain of $N=10$ nodes and restoring system 
  (\ref{rest11}), (\ref{rest13})-(\ref{rest4})
with all $\tilde \Gamma$ from list (\ref{gamma})  and the perturbation analysis 
performed for 
  the long chains of lengths $N$  from list  (\ref{long}) ($\tilde\Gamma=0.01$) yields quite similar result: parameters    $\overline{\delta}^{(T)}$ and 
$\overline{\delta}^{(K)}$ can be approximated by the lines (\ref{lines}) with the parameters  $\varkappa^{(T)}$,  $\varkappa^{(K)}$, $C^{(T)}$ and $C^{(K)}$ collected in Tables  \ref{Table:3}b  and  \ref{Table:3}c respectively.  We can rewrite system (\ref{lines}) as 
\begin{eqnarray}\label{lines2}
 \overline{\delta}^{(T)} =  10^{C^{(T)}}  \varepsilon^{\varkappa^{(T)}} , \;\;\; 
\overline{\delta}^{(K)} =  10^{C^{(K)}}\varepsilon^{ \varkappa^{(K)}}.
\end{eqnarray}
According to Table \ref{Table:3}, 
$\varkappa^{(T)}\approx  1 $ and $\varkappa^{(K)} \approx  1$ in all considered cases, i.e., both   $\overline{\delta}^{(T)}$ and  $\overline{\delta}^{(K)}$ are essentially proportional to $\varepsilon$. This fact  characterizes the robustness of the state-restoring algorithm with respect to perturbations of the restoring parameters in the Kraus operators.

{{
 \begin{table}{\footnotesize{
 \begin{tabular}{|c|cc|cc|}
 \hline
 \multicolumn{5}{|c|}{Restoring system} \cr
 \multicolumn{5}{|c|}{ (\ref{rest11}), (\ref{rest13})-(\ref{rest4}), (\ref{addconstr1}),  (\ref{addconstr2})  } \cr
 \hline
$\tilde\Gamma$ & $\varkappa^{(T)}$ & $C^{(T)}$ & $\varkappa^{(K)}$ & $C^{(K)}$ \cr
\hline
0.0001 &0.973&-0.348&0.982&0.260\cr
0.001&0.975& -0.628&0.984&-0.020\cr
0.01&0.977&-0.178&0.982& 0.262\cr
0.1&0.979&-0.096&0.981&0.251\cr
\hline
 \multicolumn{5}{c}{(a)}
 \end{tabular}}
\begin{tabular}{|c|cc|cc|}
 \hline
 \multicolumn{5}{|c|}{Restoring system} \cr
 \multicolumn{5}{|c|}{  (\ref{rest11}), (\ref{rest13})-(\ref{rest4})} \cr
 \hline
$\tilde\Gamma$ & $\varkappa^{(T)}$ & $C^{(T)}$ & $\varkappa^{(K)}$ & $C^{(K)}$ \cr
\hline
0.0001 &0.972&-0.369& 0.982& 0.261\cr
0.001&0.973& -0.652&0.984&-0.018\cr
0.01&0.979& -0.023&0.982&0.258\cr
0.1&0.977& -0.186& 0.981&0.257\cr
\hline
 \multicolumn{5}{c}{(b)}
 \end{tabular}
\begin{tabular}{|c|cc|cc|}
 \hline
 \multicolumn{5}{|c|}{Restoring system} \cr
 \multicolumn{5}{|c|}{ (\ref{rest11}), (\ref{rest13})-(\ref{rest4}), (\ref{addconstr1}), $\tilde\Gamma=0.01$} \cr
 \hline
$N$ & $\varkappa^{(T)}$ & $C^{(T)}$ & $\varkappa^{(K)}$ & $C^{(K)}$ \cr
\hline
10 &0.978&-0.123&0.983&0.208\cr
20&0.977&-0.236&0.983&   0.207\cr
30&0.980& -0.104& 0.983& 0.205\cr
40&0.980&-0.092&0.983&0.208\cr
$40_{HFST}$&0.975& -0.222&0.982&0.205\cr
\hline
 \multicolumn{5}{c}{(c)}
 \end{tabular}}
 \caption{The parameters of  linear functions (\ref{lines})  approximating the dependence of    $\log_{10} \overline{\delta}^{(T)} $  and  $\log_{10} \overline{\delta}^{(K)} $ on $\log_{10} \varepsilon$. (a) The perturbations of the chain of $N=10$ nodes with restoring system  (\ref{rest11}), (\ref{rest13})-(\ref{rest4}), (\ref{addconstr1}),  (\ref{addconstr2})  and set of different  $\tilde\Gamma$ from list (\ref{gamma}). (b)   The perturbations of the chain of $N=10$ nodes with restoring system  (\ref{rest11}), (\ref{rest13})-(\ref{rest4}) and set of different  $\tilde\Gamma$ from list (\ref{gamma}). (c) The perturbations of the chain with different $N$ from list (\ref{long}), $\tilde\Gamma=0.01$,  with restoring system  (\ref{rest11}), (\ref{rest13})-(\ref{rest4}), (\ref{addconstr1}).
All slopes $\varkappa\sim 1$ which  indicates that all approximating  lines  are essentially parallel to each other. 
 }
 \label{Table:3}
 \end{table}
}}

\section{Conclusions}
\label{Section:conclusions}

We propose a measurement-based state transfer  along the spin-1/2 chain with $XX$-interaction between spins subjected to interaction with environment via the dephasing  Lindblad operators that preserve the excitation number during the quantum evolution. We use the system of Kraus operators for state-restoring algorithm and show that this system can be generated by the unitary transformation of larger dimension. Unlike the pure state transfer along the chain non-interacting with environment studied in Ref.\cite{FWZ_arxive2025}, the presence of the Lindblad operators destroys the PST so that the transfered state  is restored in superposition with maximally mixed state with coefficient vanishing with vanishing interaction between the chain and environment. We numerically study the $\lambda$-parameters  and success probability  ($\Lambda$ and $p$) for  the 1-excitation 2-qubit state transfer and different   strengths of interaction with environment  $\tilde \Gamma$ and show that both parameters decrease with an increase in $\tilde \Gamma$. We also show that both $\Lambda$ and $p$ decrease with an increase in the  length $N$ of the homogeneous chain. However, the chain arranged for HFST yields significantly larger parameters   $\Lambda$ and $p$  which is justified by comparison of  the data in the last two column of Table \ref{Table:PST} for the chain of 40 spins.  Studying the perturbations of the restoring Kraus operators we show that the appropriate perturbation of the $T$-operator characterized by the parameter $\delta^{(T)}$  in (\ref{deltaT}) growths almost linearly with the perturbation amplitude $\varepsilon$ showing  robustness of the state-restoring algorithm with respect to perturbations of the restoring  Kraus operators.  Similar behavior is observed for the parameter $\delta^{(K)}$  
estimating the influence of the parameter perturbation   on the system of Krause operators. 

{\bf Acknowledgments.} The work was performed as a part of a state task, State Registration
No. 124013000760-0.


\section{Appendix}
\label{Section:appendix}
For the case of $k$-excitation perfect state transfer, it was shown in  Ref.\cite{FWZ_arxive2025} that $\lambda$ (it was $\lambda^2$ in notations of  Ref.\cite{FWZ_arxive2025}) is a root of the following $N^{(S)}_k$-order polynomial equation:
\begin{eqnarray}\label{lamDp}
\lambda^{ N^{(S)}_k} + \sum_{j=0}^{N^{(S)}_k-1} C_{j}(\hat V)\lambda^{j}  =0,\;\; C_j^* = C_j .
\end{eqnarray}
which can be derived from the system 
\begin{eqnarray}\label{bb1}
 \sum_{i=0}^{N^{(S)}_k-2} {\mathit{\Gamma}}_{i} (b^\dagger_j  b_i - \delta_{ij} \lambda) =b^\dagger_j b_{N^{(S)}_k-1} - \delta_{j,N^{(S)}_k-1} \lambda ,\;\; j=0,\dots,N^{(S)}_k-1 ,
\end{eqnarray}
after eliminating the parameters  ${\mathit{\Gamma}}_{i}$,
where $b_j$, $j=0,\dots,N^{(S)}_k-1$, are the columns of the  $N^{(ER)}_k \times N^{(S)}_k$ matrix $\hat V$ with the elements 
 $\hat V_{n_{ER};m_S} = V^{(k)}_{0_S0_{TL'} n_{ER}; m_S0_{TL'}0_{ER}}$. In our case of  the 1-excitation  2-qubit state transfer 
 the system (\ref{bb1})  becomes the following system of two equations:
\begin{eqnarray}
&&
b^\dagger_0 b_1 - {\mathit{\Gamma}}_0(b^\dagger_0  b_0 - \lambda)=0,\\\nonumber
&&
b^\dagger_1 b_1- \lambda - {\mathit{\Gamma}}_0 b^\dagger_1 b_0 =0.
\end{eqnarray}
 Eliminating ${\mathit{\Gamma}}_0$ we obtain the quadratic equation for $\lambda$ with roots
 \begin{eqnarray}\label{roots}
 \lambda= \frac{1}{2} \left(b_{00} + b_{11} \pm \sqrt{(b_{00} - b_{11})^2 +4 b_{01} b_{10} }\right),\;\; b_{ij} = b^\dagger_i b_j.
 \end{eqnarray}
 The minimal root  corresponds to the $\lambda$-parameter in the transfered state (\ref{protr}) with $\nu=0$.

\end{document}